\documentclass[conference]{IEEEtran}
\IEEEoverridecommandlockouts
\usepackage{cite}

\usepackage{subfigure}
\ifCLASSINFOpdf
  \usepackage[pdftex]{graphicx}
\else
  \usepackage[dvips]{graphicx}
\fi

\usepackage[cmex10]{amsmath}
\DeclareMathOperator*{\argmax}{arg\,max}
\usepackage{amssymb}
\usepackage{amsfonts}

\usepackage{algorithm}
\usepackage{algorithmic}

\usepackage{array}
\usepackage{mdwmath}
\usepackage{mdwtab}

\usepackage{url}
\newtheorem{proposition}{Proposition}
\begin{document}
\title{Radar Sensing via OTFS Signaling: A Delay Doppler Signal Processing Perspective}
\author{
  \IEEEauthorblockN{Kecheng Zhang$^{1}$, Weijie Yuan$^{1}$, Shuangyang Li$^{2}$, Fan Liu$^{1}$, Feifei Gao$^{3}$, Pingzhi Fan$^{4}$, Yunlong Cai$^{5}$}
  \IEEEauthorblockA{$^{1}$ Department of Electronic and Electrical Engineering, Southern University of Science and Technology, Shenzhen, China}
  \IEEEauthorblockA{$^{2}$ Department of Electrical, Electronic, and Computer Engineering, University of Western Australia, Perth, WA 6009, Australia}
  \IEEEauthorblockA{$^{3}$ Department of Automation, Tsinghua University, Beijing, China}
  \IEEEauthorblockA{$^{4}$ Information Coding Transmission Key Lab of Sichuan Province, CSNMT Int. Coop. Res. Centre (MoST),\\Southwest Jiaotong University, Chengdu 611756, Sichuan, China}
  \IEEEauthorblockA{$^{5}$ College of Information Science and Electronic Engineering, Zhejiang University, Hangzhou 310027, China}
  \IEEEauthorblockA{zhangkc2022@mail.sustech.edu.cn, yuanwj@sustech.edu.cn, shuangyang.li@uwa.edu.au, liuf6@sustech.edu.cn,\\ feifeigao@tsinghua.edu.cn, pzfan@swjtu.edu.cn, ylcai@zju.edu.cn}
  \thanks{This work is supported in part by National Natural Science Foundation of China under Grant 62101232, and in part by the Guangdong Provincial Natural Science Foundation under Grant 2022A1515011257.}
}
\maketitle
\begin{abstract}
  The recently proposed orthogonal time frequency space (OTFS) modulation multiplexes data symbols in the delay-Doppler (DD) domain. Since the range and velocity, which can be derived from the delay and Doppler shifts, are the parameters of interest for radar sensing, it is natural to consider implementing DD signal processing for radar sensing. In this paper, we investigate the potential connections between the OTFS and DD domain radar signal processing. Our analysis shows that the range-Doppler matrix computing process in radar sensing is exactly the demodulation of OTFS with a rectangular pulse shaping filter. Furthermore, we propose a two-dimensional (2D) correlation-based algorithm to estimate the fractional delay and Doppler parameters for radar sensing. Simulation results show that the proposed algorithm can efficiently obtain the delay and Doppler shifts associated with multiple targets.

\end{abstract}
\section{Introduction}
Future wireless systems, such as beyond 5G or 6G networks, are expected to support not only high-quality wireless communications but also highly accurate sensing services. It is widely acknowledged that sensing in the next-generation wireless network will become much more important than the currently deployed networks\cite{ISAC1}.

Most researches focus on orthogonal frequency division multiplexing (OFDM) for realizing integrated sensing and communication (ISAC) \cite{OFDM1}. OFDM has various advantages\cite{OFDM3}, such as low detection complexity, and high robustness for radar target detection. However, the implementation of OFDM could be challenging in practice\cite{OTFS0}. For example, the high peak-to-average power ratio (PAPR) problem of OFDM will reduce power efficiency, especially at high-frequency carriers. Moreover, in high-mobility environments, the orthogonality between subcarriers is destroyed due to the severe Doppler effect, which could degrade communication performance significantly. Besides, the channel response in a high-mobility communication scenario varies significantly across different coherence regions, in which OFDM needs to estimate the channel more frequently to get accurate channel state information and leading to high signaling overhead. These problems motivate the researchers to develop a new modulation technique that has robust communication performance in high-mobility environments.

The recently proposed orthogonal time frequency space (OTFS) modulation \cite{OTFS0}, which has the same waveform as the vector-OFDM (VOFDM) proposed in \cite{VOFDM}, has become a promising candidate for providing reliable communication under high-mobility multi-path scenarios\cite{AndrewZhang}. Different from the conventional time-frequency (TF) domain modulation schemes, the OTFS modulation scheme multiplexes information symbols over the two-dimensional (2D) delay-Doppler (DD) domain\cite{OTFS0}, where the resolvable paths of the wireless channel are characterized by the different delay and Doppler shifts. There are many benefits to describing the channel in the DD domain. For example, in high-mobility communication scenarios, the DD domain channel is generally sparse and quasi-static\cite{AndrewZhang} compared to the fast time-varying channel in the TF domain. Meanwhile, OTFS enjoys a lower PAPR\cite{Low_PAPR}, making it easier to be implemented with a more efficient power amplifier, compared to the OFDM counterpart. More importantly, the range and velocity parameters of the targets, which can be inferred from delay and Doppler shifts, are the primary parameters to be estimated in radar signal processing, making OTFS a natural choice for realizing ISAC.

Some recent works have considered radar sensing via OTFS\cite{WORK1, WORK2}. In \cite{WORK1}, the author proposed a maximum likelihood (ML) algorithm to estimate range and velocity via the OTFS transmission scheme. It shows that the OTFS signal can achieve radar estimation performance bound while maintaining a superior communication performance over OFDM. A matched-filter algorithm was proposed in \cite{WORK2} to estimate the range and velocity of targets, in which the structure of the effective OTFS channel is utilized to simplify the computation. It is shown that the estimation performance of target speed based on OTFS is better than using OFDM.

However, the potential of radar sensing using OTFS has not been fully explored. For example, the complexity of the algorithm in \cite{WORK1} is nearly the cube of the OTFS frame size, which is relatively high. The work of \cite{WORK2} only considers the integer delay and Doppler shifts. However, this could be highly impractical in real wireless networks, where the time resource is limited, resulting in insufficient Doppler resolution. Consequently, the presence of fractional Doppler is inevitable. Moreover, for an accurate sensing performance, it is important to study radar sensing with insufficient frequency resources, despite the fact that the integer delay is generally sufficient for communication design. Thus, it is necessary to consider both the fractional delay and Doppler shifts for radar sensing. The aforementioned issues motivate us to develop a  method to fulfill radar sensing via OTFS signaling in the presence of fractional delay and Doppler shifts.

In this paper, we first discuss the intrinsic connection between the radar sensing range-Doppler matrix computation and the OTFS demodulation process that they are exactly the same under a rectangular pulse shaping filter. Then, inspired by the pulse compression (fast-time matched filtering) in radar sensing\cite{RADAR_AND_OTFS0}, we propose a two-step method to estimate the fractional delay and Doppler indices. In particular, the sensing receiver acquires the DD domain echo waves reflected by the targets and performs a 2D correlation between the received DD domain symbols and the transmitted information symbols. After getting the correlation matrix, a difference-based method, in which we take the subtraction between the indices of the second and first maximum magnitude of the correlated matrix, is implemented to calculate the fractional delay and Doppler indices. Simulation results show that the proposed algorithm can obtain the delay and Doppler shifts associated with multiple targets efficiently.

\textit{Notations:} $(\cdot)^{*}$ denotes the conjugate operation; $[\cdot]_{N}$ denotes the modulo operation with respect to (w.r.t.) $N$; $|\cdot|$ means taking the magnitude of a complex; $\delta(\cdot)$ is the Dirac delta function; the bold lowercase $\mathbf{a}^{N}$ represents a vector with N dimension, and the bold uppercase $\mathbf{A}^{M\times N}$ represents a matrix with $M\times N$ dimension; $\mathbb{R}^{M\times N}$, $\mathbb{Z}^{M\times N}$, and $\mathbb{C}^{M\times N}$ are the $M\times N$ dimension matrix spaces with real, integer, and complex entries, respectively.
\section{System Model}\label{model}
For each OTFS frame, the number of time slots and the number of sub-carriers are denoted by $N$ and $M$, respectively. The occupied bandwidth of one OTFS frame is $M\Delta f$ with a duration $NT$, where $\Delta f$ represents the subcarrier space, and $T$ is the symbol duration. The sequence of information bits is mapped to a symbol set $\{X_{\text{DD}}[k,l],k=0,\dots,N-1,l=0,\dots,M-1\}$ in the DD domain, where \textit{l} and \textit{k} represent the indices of delay and Doppler shifts, respectively, and $X_{\text{DD}}\in \mathbb{A}$, in which $\mathbb{A}=\{a_{1},\dots,a_{|\mathbb{A}}|\}$ is the modulation alphabet (e.g. QAM). The DD domain symbol $X_{\text{DD}}[k,l]$ is transformed into the TF domain signal $X_{\text{TF}}[n,m]$ through the inverse symplectic finite Fourier transform (ISFFT)\cite{OTFS0},
\begin{equation}\label{eq1}
  X_{\text{TF}}[n,m]=\frac{1}{\sqrt[]{NM}}\sum_{k = 0}^{N-1} \sum_{l=0}^{M-1}X_{\text{DD}}[k,l]e^{j2\pi(\frac{nk}{N}-\frac{ml}{M})}\text{.}
\end{equation}
The TF domain modulator maps $X_{\text{TF}}[n,m]$ to the time domain transmit signal $s(t)$ via the Heisenberg transform\cite{OTFS0},
\begin{equation}\label{eq2}
  s(t)=\sum_{n=0}^{N-1}\sum_{m=0}^{M-1}X_{\text{TF}}[n,m]g_{tx}(t-nT)e^{j2\pi m\Delta f(t-nT)}\text{,}
\end{equation}
where $g_{tx}(t)$ is the pulse shaping filter at the transmitter side.

The signal $s(t)$ transmits over a linear time-varying wireless channel and is reflected by the sensing targets, yielding the sensing echo as
\begin{equation}\label{eq3}
  r(t)=\int\int h(\tau,\nu)e^{j2\pi\nu(t-\tau)}s(t-\tau) \,d\tau\,d\nu+z(t)\text{,}
\end{equation}
where $z(t)$ denotes the additive white Gaussian noise (AWGN) process with one side power spectral density (PSD) $N_0$, and $h(\tau,\nu)\in \mathbb{C}$ is the complex base-band channel impulse response in the DD domain, which can be expressed as
\begin{equation}\label{eq4}
  h(\tau,\nu)=\sum_{i=1}^{P}h_{i}\delta(\tau-\tau_{i})\delta(\nu-\nu_{i})\text{,}
\end{equation}
where $P\in \mathbb{Z}$ is the number of targets in the sensing scenario, and $h_{i}$, $\tau_{i}$ and $\nu_{i}$ denote the reflection coefficient, delay, and Doppler shift associated with the $i$th target, respectively. Assuming that the range and relative velocity associated with the $i$th target are $R_{i}$ and $V_{i}$, respectively, the round-trip delay $\tau_{i}\in\mathbb{R}$ and the Doppler frequency $\nu_{i}\in\mathbb{R}$ are expressed as
\begin{equation}\label{eq56}
  \tau_{i}=\frac{2R_{i}}{c}=\frac{l_{\tau_{i}}}{M\Delta f}\text{, }
  \nu_{i}=\frac{2f_{c}V_{i}}{c}=\frac{k_{\nu_{i}}}{NT}\text{,}
\end{equation}
where $c$ is the speed of light and $f_{c}$ is the carrier frequency, $l_{\tau_{i}}=l_{i}+\iota_{i}$ and $k_{\nu_{i}}=k_{i}+\kappa_{i}$ are the delay and Doppler indices of the $i$th target, $l_{i}\in\mathbb{Z}$ and $k_{i}\in\mathbb{Z}$ denote the integer parts of indices at the $i$th target, while $\iota_{i}\in[-0.5,0.5]$ and $\kappa_{i}\in[-0.5,0.5]$ denote the fractional parts.

At the OTFS receiver, the received time domain signal $r(t)$ is first transformed into the TF domain via the Wigner transform\cite{OTFS0}, which is given by
\begin{equation}\label{eq7}
  Y_{\text{TF}}[n,m]=\int_{-\infty}^{\infty}r(t)g^{*}_{rx}(t-nT)e^{-j2\pi m\Delta f(t-nT)}\,dt\text{,}
\end{equation}
where $g_{rx}(t)$ is the receiver pulse shaping filter. Then the TF domain received signal $Y_{\text{TF}}[n,m]$ is transformed onto the DD domain through the SFFT, which can be expressed as
\begin{equation}\label{eq8}
  Y_{\text{DD}}[k,l]=\frac{1}{\sqrt[]{NM}}\sum_{n=0}^{N-1}\sum_{m=0}^{M-1}Y_{\text{TF}}[n,m]e^{-j2\pi(\frac{nk}{N}-\frac{ml}{M})}\text{.}
\end{equation}
According to \cite{OTFS_Window_Design}, the DD domain input-output relationship in the delay-Doppler domain can be written as
\begin{equation}\label{eq9}
  \begin{aligned}
    Y_{\text{DD}}[k,l] & =\sum_{i=1}^{P}h_{i}e^{j2\pi\frac{(l-l_{\tau_{i}})k_{\nu_{i}}}{MN}}\alpha(k-k_{\nu_{i}},l-l_{\tau_{i}}) \\
                       & \times X_{\text{DD}}[[k-k_{\nu_{i}}]_{N},[l-l_{\tau_{i}}]_{M}]+Z_{\text{DD}}[k,l]\text{,}
  \end{aligned}
\end{equation}
where $Z_{\text{DD}}[k,l]$ denotes the effective noise in the DD domain that follows the Gaussian distribution $\mathcal{CN}(0, \sigma^2)$, and $\alpha_{i}(k,l)$ is the phase offset given by
\begin{equation}\label{phase_offset}
  \alpha(k,l)=\left\{
  \begin{aligned}
     & 1,                     & l\geqslant 0, \\
     & e^{-j2\pi\frac{k}{N}}, & l<0\text{.}
  \end{aligned}
  \right.
\end{equation}
\section{OTFS Based Radar Sensing}\label{proposed}
In this section, we first show the intrinsic connection between the range-Doppler matrix computation in radar sensing and the OTFS demodulation. Then we propose a 2D correlation-based method to estimate the delay and Doppler indices.
\subsection{Range-Doppler Matrix Computation via OTFS Demodulation}
\begin{figure}[t]
  \centering
  \includegraphics[width=0.8\columnwidth]{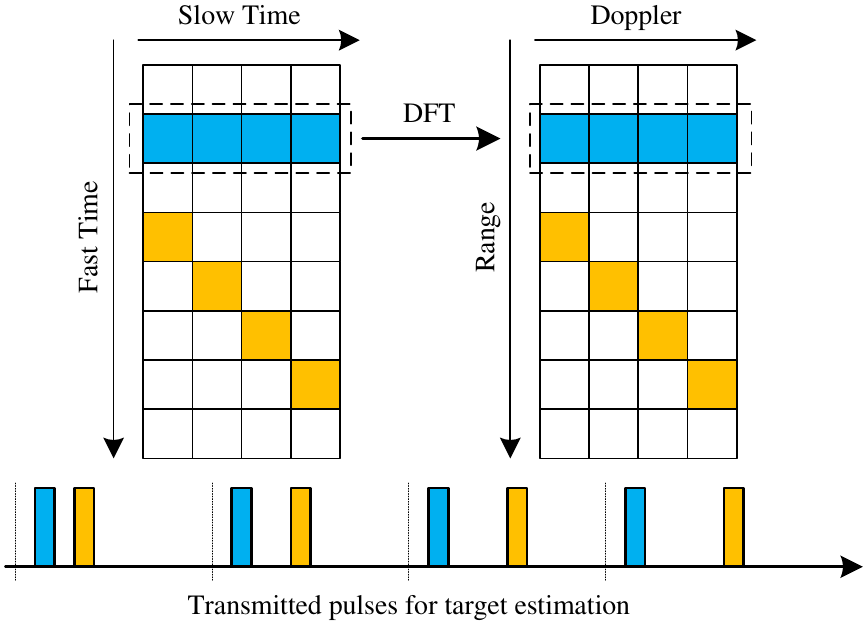}
  \caption{An illustration for obtaining the fast-time slow-time matrix and the range-Doppler matrix.}
  \label{DFT_RD}
\end{figure}
In radar sensing, we need several pulses to get the range/Doppler information. The sampling rate of every transmitted pulse is called fast time, and the sampling interval between the pulses is called slow time\cite{RADAR_AND_OTFS0}. Fig. \ref{DFT_RD} is an example to show how to calculate the range-Doppler matrix. There are four pulses for target estimation in this example. By rearranging the samples of the pulses, we get the fast-time slow-time matrix. The range-Doppler matrix can then be obtained by applying the discrete Fourier transform (DFT) along the slow-time axis. By denoting $y_{\text{TD}}[n]=r(\frac{nT}{M})$, we can see that the fast-time slow-time matrix $\mathbf{R}^{MN}$ is formulated by assigning $y_{\text{TD}}[m+nM]$ to the $(m,n)$-th entry of $\mathbf{R}$. Now combine (\ref{eq7}) and (\ref{eq8}), the demodulation procedure can be expressed as below
\begin{equation}\label{zak_transform}
  Y_{\text{DD}}[k,l]=\frac{1}{\sqrt[]{N}}\sum_{n=0}^{N-1}y_{\text{TD}}[l+nM]e^{-j2\pi\frac{nk}{N}}\text{,}
\end{equation}
which is the discrete Zak transform (DZT)\cite{OTFS0} under the energy-normalized rectangular matched filter. It can be observed that the DFT operation on each row of the 2D fast-time slot-time matrix is exactly the DZT in (\ref{zak_transform}), which means the matrix $\mathbf{Y}_{\text{DD}}$ is the same as the range-Doppler matrix in radar sensing if we choose the rectangular pulse shaping filter.
\subsection{2D Correlation-based Parameter Estimator}
We will describe the 2D correlation-based parameter estimator and the fractional delay and Doppler calculating method in this subsection. According to the brief description in the previous subsection, the received DD domain symbol matrix $\mathbf{Y}_{\text{DD}}$ is the same as the range-Doppler matrix. However, we cannot localize the targets of interest from $\mathbf{Y}_{\text{DD}}$ directly, due to the presence of information symbols. After passing the channel, the delay and Doppler bins of these DD domain symbols will overlap with each other, which makes the received DD domain signals contain both the channel responses and the overlapped responses from DD domain information symbols. Inspired by pulse compression in radar sensing, a 2D correlation-based estimator, which can be considered as pulse compression along both the delay and Doppler axes, is implemented to improve the acquisition of delay and Doppler parameters.

Denote the matrix after 2D pulse compression as $\mathbf{V}$, then the accumulated correlation coefficient under different delay and Doppler indices can be expressed as
\begin{align}\label{correlation}
  V[k,l]=\sum_{n=0}^{N-1}\sum_{m=0}^{M-1} & Y^{*}_{\text{DD}}[n,m]X_{\text{DD}}[[n-k]_{N},[m-l]_{M}]\nonumber \\
                                          & \times\alpha[n-k,m-l]e^{j2\pi\frac{(m-l)k}{NM}}\text{,}
\end{align}
where $k\in[0,N-1]$ and $l\in[0,M-1]$, and $\alpha[k,l]$ is a phase offset given in (\ref{phase_offset}).
It should be noted that the 2D correlation cannot be considered as a simple combination of two one-dimensional correlation operations, since there is a phase term $\alpha[n-k,m-l]$ in (\ref{correlation}).
\begin{figure} \centering
  \subfigure[Before performing the 2D correlation.] {\label{received_matrix}
    \includegraphics[width=0.4655\columnwidth]{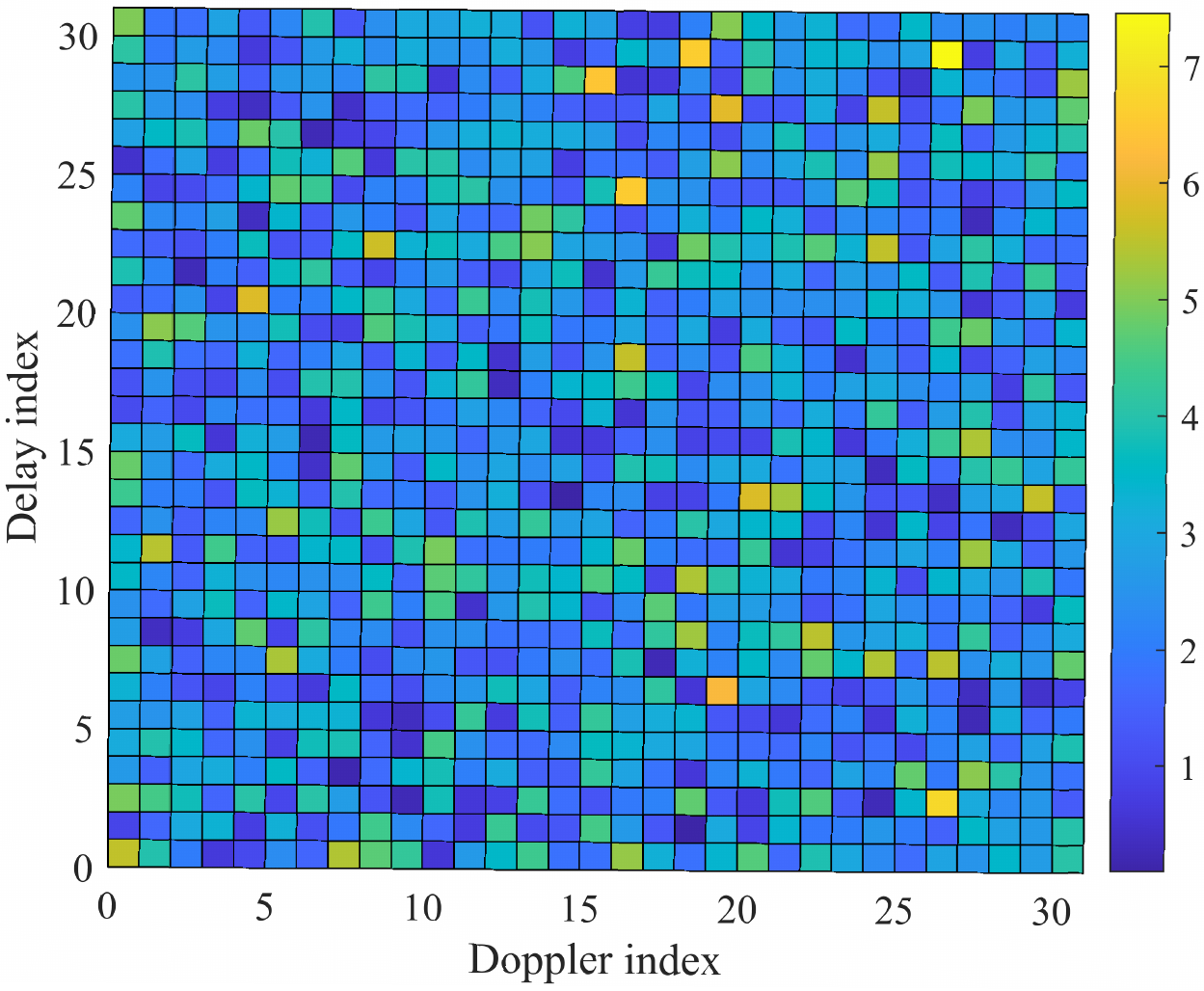}
  }
  \subfigure[After performing the 2D correlation.] { \label{correlation_fig}
    \includegraphics[width=0.465\columnwidth]{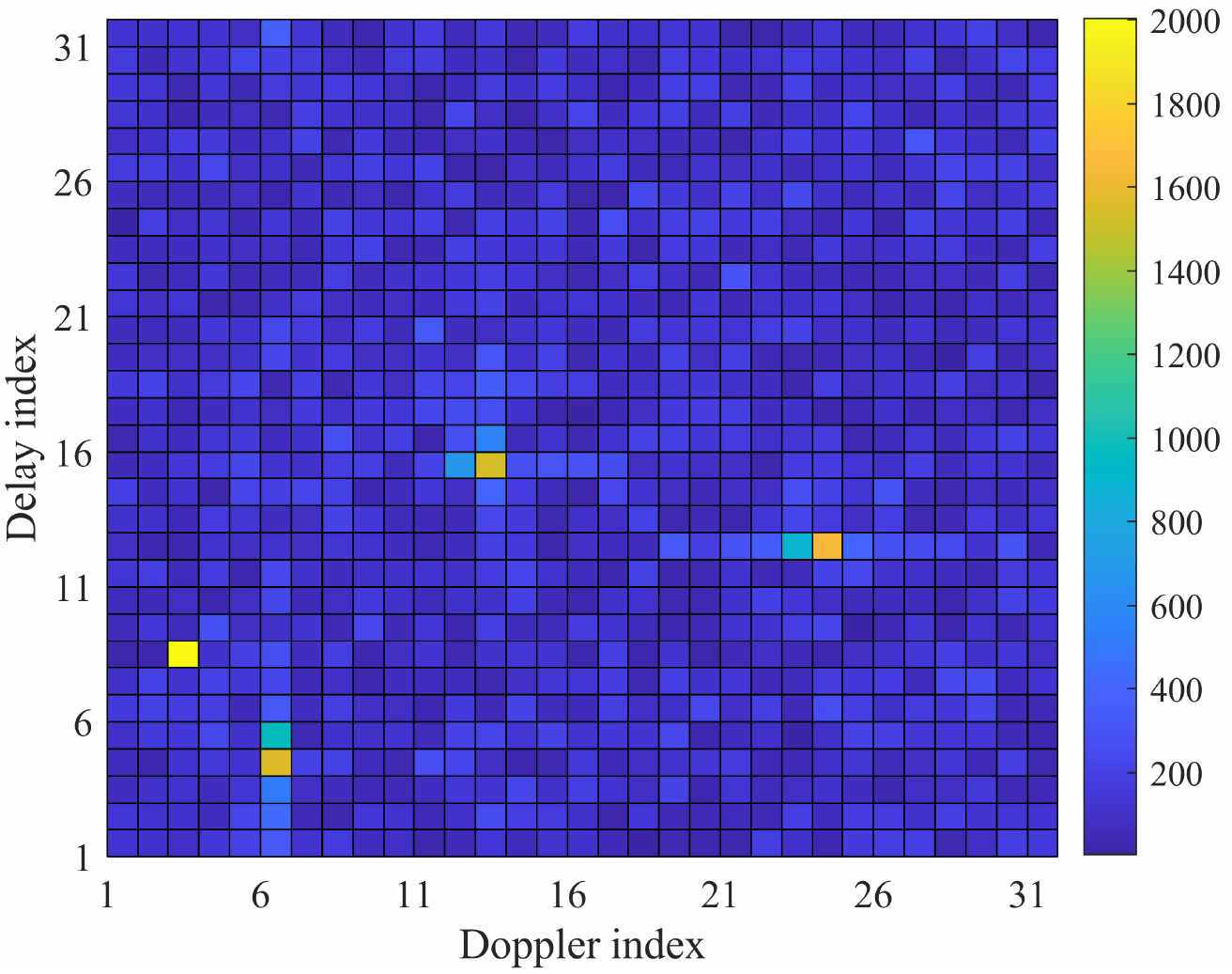}
  }
  \caption{The delay-Doppler matrices for radar sensing via OTFS waveform before and after performing the 2D correlation, where $P=4$ targets are considered and the delay and Doppler indices are of fractional values}
  \label{matrix_figure}
\end{figure}

To have a better illustration of the proposed method, we establish an example of OTFS radar sensing under the noiseless scenario. In this example, we assume there are $P=4$ targets, and the normalized quadrature phase shift keying (QPSK) information symbols are generated randomly. We set $M=32$ and $N=32$, and the delay and Doppler indices associated with different targets are set as $[24.25, 18.07, 11.72, 21.30]$ and $[2.58,3.93,2.04,-3.24]$, respectively. The DD domain received symbol matrix under this specific scenario is represented in Fig. \ref{received_matrix}, and the matrix after pulse compression is shown in Fig. \ref{correlation_fig}. We can see that the received symbol matrix $\mathbf{Y}_{\text{DD}}$, i.e., the range-Doppler matrix in radar sensing, is dense due to the overlapped responses from each DD domain information symbol, as described in (\ref{eq9}). But after performing the 2D correlation operation, the target responses are more localized. This procedure can be viewed as a special pulse compression in the DD domain, which has a similar function to pulse compression in radar sensing to enhance the acquisition of delay and Doppler responses.

After obtaining the 2D correlation matrix, we can estimate the delay and Doppler indices by taking the matrix peaks. By finding the peaks of the range-Doppler matrix, we can only get the integer parts of the indices, while the fractional parts of the delay and Doppler indices remain unknown. This will lead to an inaccurate sensing result\cite{OTFS3}. By observing the peaks in the 2D correlation matrix, we can see that there is power leakage from the corresponding delay-Doppler bins to their neighbors, as shown in Fig. \ref{correlation_fig}. The power leakage is caused by the presence of fractional delay and Doppler indices. Inspired by this obsession, we propose a simple method to estimate the fractional delay and Doppler through a difference method, which is explained as follows.

We first consider how to calculate the fractional part of the Doppler index under the noiseless scenario. Suppose that the delay and Doppler indices of one target are $k_{\nu_{i}}=k_{i}+\kappa_{\nu_{i}}$ and $l_{\tau_{i}}=l_{i}+\iota_{\tau_{i}}$, respectively. Meanwhile, assume that there are no two paths that have the same delay. Then we can obtain the row index of the maximum magnitude $k_{\nu_{1}}^{\prime}$ and the row index of the second maximum magnitude $k_{\nu_{2}}^{\prime}$ at the $l_{i}$th column in the delay-Doppler matrix after pulse compression,
\begin{equation}\label{get_two_max}
  \begin{aligned}
    k_{\nu_{1}}^{\prime} & =\argmax_{k\in\left\{\left\lceil -N/2 \right\rceil,\dots,\left\lceil N/2 \right\rceil-1\right\}}|V[k,l_{i}]|\text{,}                          \\
    k_{\nu_{2}}^{\prime} & =\argmax_{k\in\left\{\left\lceil -N/2 \right\rceil,\dots,\left\lceil N/2 \right\rceil-1\right\}\backslash\{k_{\nu_{1}}^{\prime}\}}|V[k,l_{i}]|\text{.}
  \end{aligned}
\end{equation}
Having $k_{\nu_{1}}$ and $k_{\nu_{2}}$ in hand, we have the following proposition.
\begin{proposition}
  Under noiseless conditions, the actual Doppler index $k_{i}+\kappa_{\nu_{i}}$ must fall into the interval bounded by $k_{\nu_{1}}$ and $k_{\nu_{2}}$, where $|k_{\nu_{2}}^{\prime}-k_{\nu_{1}}^{\prime}|=1$. Then, the ratio between the magnitudes of the correlation coefficients with the same delay index and Doppler indices $k_{\nu_{1}}$ and $k_{\nu_{2}}$ can be approximated by
  \begin{equation}\label{derivation}
    \frac{|V[k_{\nu_{1}}^{\prime},l_{i}]|}{|V[k_{\nu_{2}}^{\prime},l_{i}]|}\approx\frac{|k_{\nu_{2}}^{\prime}-k_{\nu_{1}}^{\prime}-\kappa_{\nu_{i}}|}{|-\kappa_{\nu_{i}}|}\text{,}
  \end{equation}
  where the approximation error is at the order of $\mathcal{O}\left(\frac{1}{MN}\right)$.
\end{proposition}
\begin{proof}
  See Appendix \ref{proofa}.
\end{proof}
Therefore, the fractional Doppler in (\ref{eq56}) can be derived as
\begin{equation}\label{get_kappa}
  \kappa_{i}=\frac{(k_{\nu_{2}}^{\prime}-k_{\nu_{1}}^{\prime})|V[k_{\nu_{2}}^{\prime},l_{i}]|}{|V[k_{\nu_{1}}^{\prime},l_{i}]|+|V[k_{\nu_{2}}^{\prime},l_{i}]|}\text{,}
\end{equation}
Similarly, by applying the above derivation to the fractional delay taps, we can get
\begin{equation}\label{get_iota}
  \iota_{i}=\frac{(l_{\tau_{2}}^{\prime}-l_{\tau_{1}}^{\prime})|V[k_{i},l_{\tau_{2}}^{\prime}]|}{|V[k_{i},l_{\tau_{1}}^{\prime}]|+|V[k_{i},l_{\tau_{2}}^{\prime}]|}\text{,}
\end{equation}
where $l_{\tau_{1}}^{\prime}$ and $l_{\tau_{2}}^{\prime}$ are the column indices of the maximum and the second maximum magnitudes in the $k_{i}$th row of the 2D correlation matrix.

The algorithm to estimate the fractional parts of the delay and Doppler indices are summarized in Algorithm \ref{estimate_algorithm}, where $\boldsymbol{k}=[k_{1},\dots,k_{P}]$ and $\boldsymbol{l}=[l_{1},\dots,l_{P}]$ denotes the integer parts of the Doppler and delay indices, respectively. By assuming that the number of targets $P$ is known, this algorithm has a complexity of $\mathcal{O}(M^2N^2)$. But a more realistic approach is to determine whether the target exists or not by setting a detection threshold for the 2D correlation matrix, which will be our work in the future. Now we briefly discuss a possible solution for peak selection.
\begin{figure}[t]
  \centering
  \includegraphics[width=0.75\columnwidth]{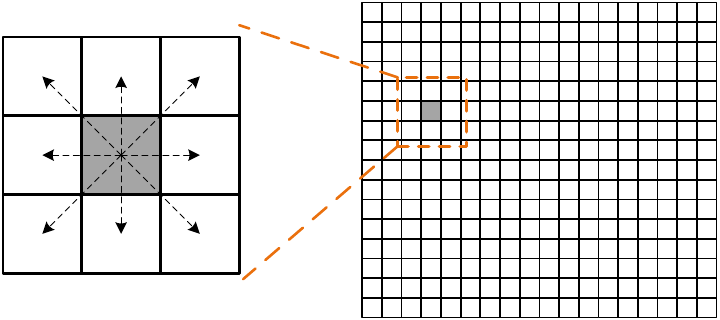}
  \caption{The illustration concerning how to pick the peaks in $\mathbf{V}$}
  \label{pick_peak}
\end{figure}

As shown in Fig.\ref{pick_peak}, we first select one entry and compare the magnitude of this entry with its eight adjacent entries. If its magnitude is larger than all eight neighbors, then this entry is considered one peak. Apply this method over the whole matrix, we get all the peaks that appeared in $\mathbf{V}$. Then we pick the largest $P$ peaks and consider the corresponding row and column indices are the integer parts of delay and Doppler shifts denoted as $\boldsymbol{k}$ and $\boldsymbol{l}$. By implementing Algorithm \ref{estimate_algorithm}, we can estimate the delay and Doppler shifts associated with the off-grid targets.
\begin{algorithm}
  \caption{Estimate the fractional delay and Doppler indices from 2D correlated delay-Doppler matrix}
  \label{estimate_algorithm}
  \begin{algorithmic}[1]
    \REQUIRE $\mathbf{Y}_{\text{DD}},\mathbf{X}_{\text{DD}},M,N,P$
    \ENSURE $\boldsymbol{\kappa_{\nu}}$, $\boldsymbol{\iota_{\tau}}$
    \STATE Compute $\mathbf{V}$ via (\ref{eq9}) and (\ref{correlation}).
    \STATE Pick the largest $P$ peaks in $\mathbf{V}$, get $\boldsymbol{k}$ and $\boldsymbol{l}$
    \FOR{$i=1; i\leqslant P$}
    \STATE $k_{i}\gets k_{\nu_{1}}$
    \IF {$|V[[k_{i}+1]_{N}, l_{i}]|>|V[[k_{i}-1]_{N}, l_{i}]|$}
    \STATE $k_{\nu_{2}}\gets[k_{i}+1]_{N}$
    \ELSE
    \STATE $k_{\nu_{2}}\gets[k_{i}-1]_{N}$
    \ENDIF
    \STATE Calculate $\kappa_{\nu_{i}}$ using (\ref{get_kappa})
    \STATE $l_{i}\gets l_{\tau_{1}}$
    \IF {$|V[k_{i},[l_{i}+1]_{N}]|>|V[k_{i}, [l_{i}-1]_{M}]|$}
    \STATE $l_{\tau_{2}}\gets[l_{i}+1]_{M}$
    \ELSE
    \STATE $l_{\tau_{2}}\gets[l_{i}-1]_{M}$
    \ENDIF
    \STATE Calculate $\iota_{\tau_{i}}$ using (\ref{get_iota})
    \ENDFOR
  \end{algorithmic}
\end{algorithm}
\section{Numerical Results}\label{result_analysis}
In this section, we investigate the estimation performance under various conditions through Monte Carlo simulations. All simulation results are averaged from $10^4$ OTFS frames. We set $M=128$ and $N=64$ for each OTFS frame, which means there are 64 time slots and 128 subcarriers in the TF domain. The information bits are generated randomly and mapped to QPSK symbols. The carrier frequency is set as 24 GHz with 39 kHz subcarrier spacing. The maximum speed of the mobile user is set to be 440 km/h. The OTFS simulation performance is compared with the conventional periodogram-based OFDM radar sensing algorithm, where the simulation parameters are the same.

\begin{figure} \centering
  \subfigure[The RMSE of velocity estimation with respec to SNR.] {\label{velocity_result}
    \includegraphics[width=0.465\columnwidth]{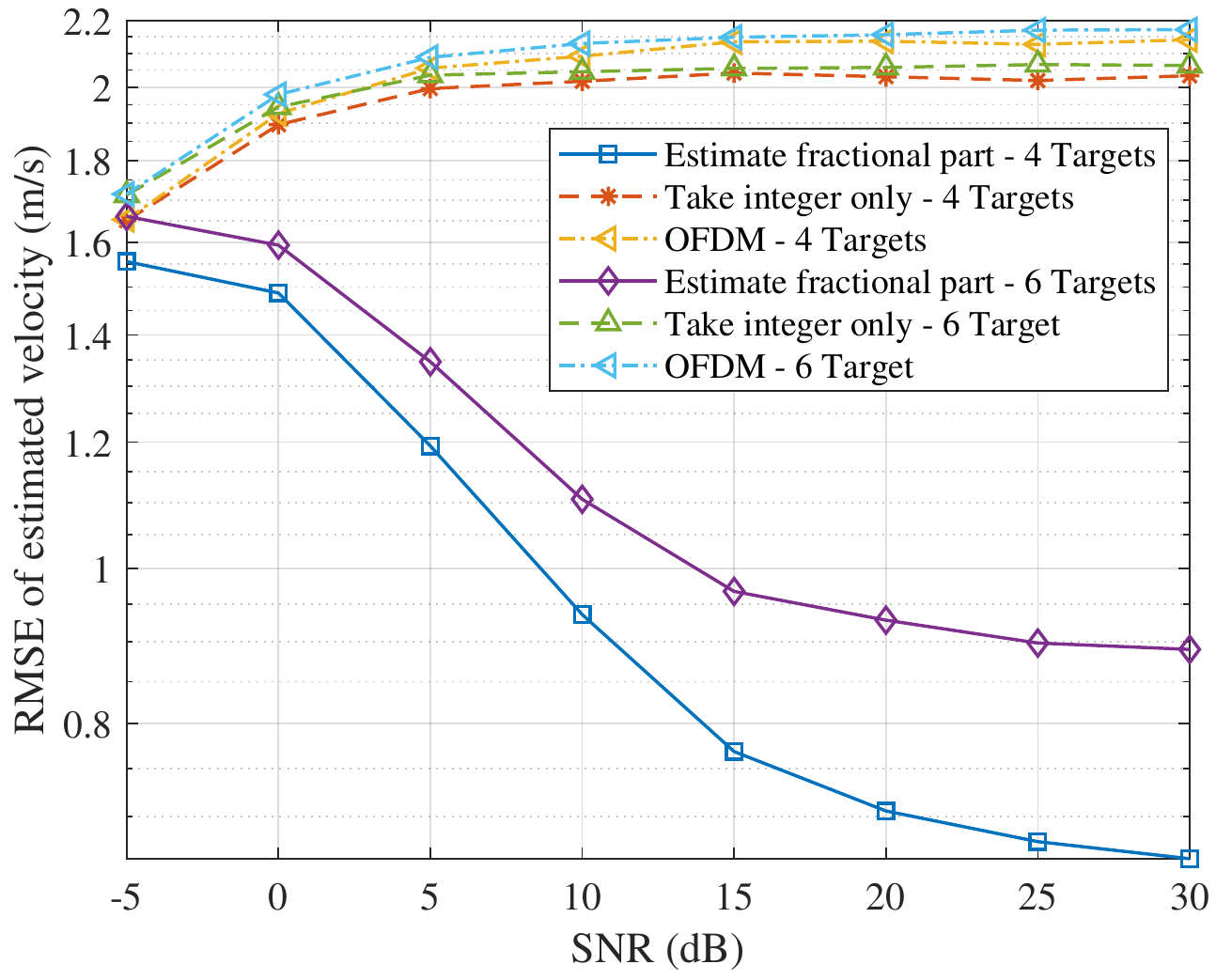}
  }
  \subfigure[The RMSE of range estimation with respec to SNR.] { \label{range_result}
    \includegraphics[width=0.465\columnwidth]{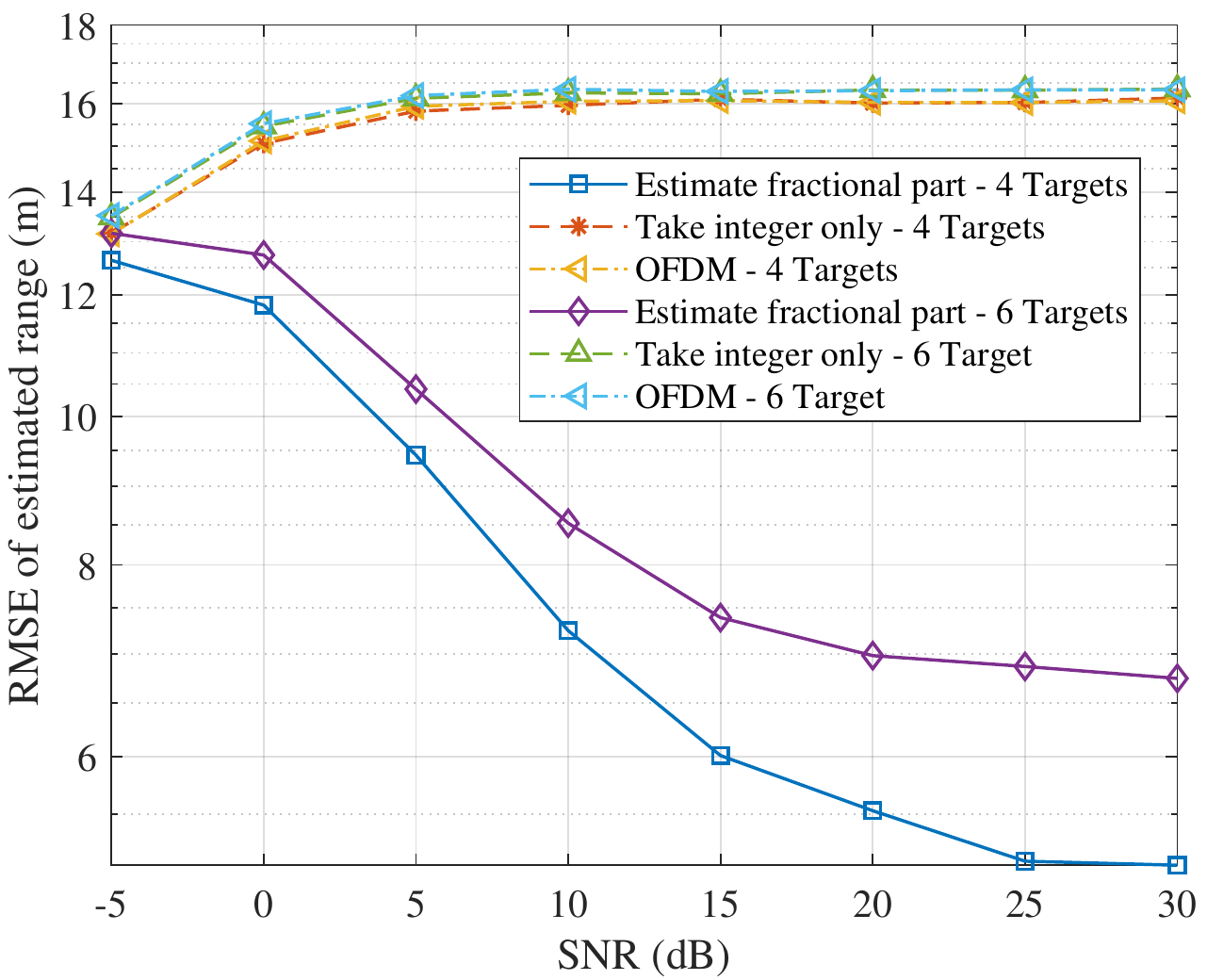}
  }
  \caption{The estimation performance versus different SNR.}
  \label{est_performance}
  \vspace{-3mm}
\end{figure}
In the simulations, we consider two scenarios with $P=4$ and $P=6$ targets and calculate the root-mean-square error (RMSE) of estimated range and velocity versus signal-to-noise ratio (SNR). The RMSE of the proposed parameter estimation algorithm is compared with the methods that only take the integer part of the delay and Doppler taps from the 2D correlation matrix. The simulation results are shown in Fig. \ref{velocity_result} and Fig. \ref{range_result}.

It is observed in Fig. \ref{velocity_result} that when there are 4 targets, the RMSE of the estimated velocity is much lower if the fractional Doppler index is estimated through (\ref{get_kappa}), which shows the effectiveness of the proposed algorithm. When there are 6 targets, the RMSE performance of the proposed algorithm becomes worse at the same SNR level. This is because the interference caused by the response overlaps from each symbol is more severe if there are 6 targets, which makes the approximation in (\ref{derivation}) less accurate. Thus, the velocity estimation becomes worse. For the same reason, the RMSE of the estimated range, shown in Fig.\ref{range_result}, is smaller if there are 4 targets in the simulation scenario, and the estimation results are much worse if only the integer part of the indices is considered.

If we only take the integer parts of the parameters, it can be observed that the RMSE of range estimation is hardly changed versus SNR. This is because the range and velocity resolutions of each grid are $\frac{c_0}{M\Delta f}=30$ m and $\frac{c_0}{NTf_c}=3.8$ m/s, respectively, where $c_0$ is lightspeed and $f_c$ the carrier frequency. The average minimum errors are half of the resolution, which accords with the simulation results. Meanwhile, the estimation performance of OFDM sensing is the same as the OTFS sensing if we only take the integer parts.
\begin{figure*}[h]
  \begin{align*}\label{v2_complicated}
    &\mathbb{E}[V[k,l]^2]=\mathbb{E}\left\{\left[\sum_{n^{\prime},m^{\prime},k^{\prime},l^{\prime}}
    h^{*}_{\omega}[n^{\prime}-k^{\prime}, m^{\prime}-l^{\prime}]X_{\text{DD}}^{*}[k^{\prime},l^{\prime}]X_{\text{DD}}[n^{\prime}-k,m^{\prime}-l]
    +\sum_{n^{\prime},m^{\prime}}
    X_{\text{DD}}[n^{\prime}-k,m^{\prime}-l]Z_{\text{DD}}^{*}[n^{\prime},m^{\prime}]\right]\right.\\
    &\left.\times\left[\sum_{n^{\prime\prime},m^{\prime\prime},k^{\prime\prime},l^{\prime\prime}}
    h^{*}_{\omega}[n^{\prime\prime}-k^{\prime\prime}, m^{\prime\prime}-l^{\prime\prime}]X_{\text{DD}}^{*}[k^{\prime\prime},l^{\prime\prime}]X_{\text{DD}}[n^{\prime\prime}-k,m^{\prime\prime}-l]
    +\sum_{n^{\prime\prime},m^{\prime\prime}}
    X_{\text{DD}}[n^{\prime\prime}-k,m^{\prime\prime}-l]Z_{\text{DD}}^{*}[n^{\prime\prime},m^{\prime\prime}]\right]\right\}\\
&=\sum_{n^{\prime},m^{\prime},k^{\prime},l^{\prime},n^{\prime\prime},m^{\prime\prime},k^{\prime\prime},l^{\prime\prime}}
h^{*}_{\omega}[n^{\prime}-k^{\prime}, m^{\prime}-l^{\prime}]h^{*}_{\omega}[n^{\prime\prime}-k^{\prime\prime}, m^{\prime\prime}-l^{\prime\prime}]\mathbb{E}\{X_{\text{DD}}^{*}[k^{\prime},l^{\prime}]X_{\text{DD}}[n^{\prime}-k,m^{\prime}-l]X_{\text{DD}}^{*}[k^{\prime\prime},l^{\prime\prime}]\\
&\times X_{\text{DD}}[n^{\prime\prime}-k,m^{\prime\prime}-l]\}+\sum_{n^{\prime},m^{\prime},k^{\prime},l^{\prime},n^{\prime\prime},m^{\prime\prime}}h^{*}_{\omega}[n^{\prime}-k^{\prime}, m^{\prime}-l^{\prime}]\mathbb{E}\{X_{\text{DD}}^{*}[k^{\prime},l^{\prime}]X_{\text{DD}}[n^{\prime}-k,m^{\prime}-l]X_{\text{DD}}[n^{\prime\prime}-k,m^{\prime\prime}-l]\\
&\times Z_{\text{DD}}^{*}[n^{\prime\prime},m^{\prime\prime}]\}+\sum_{n^{\prime\prime},m^{\prime\prime},k^{\prime\prime},l^{\prime\prime},n^{\prime},m^{\prime}}h^{*}_{\omega}[n^{\prime\prime}-k^{\prime\prime}, m^{\prime\prime}-l^{\prime\prime}]\mathbb{E}\{X_{\text{DD}}^{*}[k^{\prime\prime},l^{\prime\prime}]X_{\text{DD}}[n^{\prime}-k,m^{\prime}-l]X_{\text{DD}}[n^{\prime\prime}-k,m^{\prime\prime}-l]\\
    &\times Z_{\text{DD}}^{*}[n^{\prime},m^{\prime}]\}+\sum_{n^{\prime},m^{\prime},n^{\prime\prime},m^{\prime\prime}}\mathbb{E}\{X_{\text{DD}}[n^{\prime}-k,m^{\prime}-l]X_{\text{DD}}[n^{\prime\prime}-k,m^{\prime\prime}-l]Z_{\text{DD}}^{*}[n^{\prime\prime},m^{\prime\prime}]Z_{\text{DD}}^{*}[n^{\prime},m^{\prime}]\}.\tag{21}
  \end{align*}
  \hrule
  \vspace{-3mm}
\end{figure*}
\section{Conclusion}\label{conclusion}
This paper unveiled the intrinsic connection between the demodulation procedure of OTFS signaling and the range-Doppler matrix computation in radar sensing and applied a 2D correlation-based method to estimate the delay and Doppler indices. Since the delay and Doppler indices of the channel are usually fractional in the DD domain for off-grid targets, we proposed a difference-based method to estimate the fractional part of the parameters. Simulation results showed that the proposed algorithm can obtain the range and velocity estimates corresponding to the off-grid targets accurately.
\begin{appendices}
  \section{Proof of Proposition 1}\label{proofa}
  For ease of exposition, we only consider the ideal pulse shaping filter in the proof. As shown in \cite{raviteja2018interference}, the non-zero entries in the DD domain effective channel are localized identically under both ideal and rectangular pulse shaping filters. The entries at the same location differs with only a particular phase offset. Therefore, our derived results can be extended to the case under rectangular pulse straightforwardly.

Let us rewrite the input-output relationship in (8) as a 2D convolution form \cite{OTFS_Window_Design}, given by\footnote{For brevity, we will omit the lower and upper limits of the summation operator in what follows. In particular, the indices $n$, $n^{\prime}$, $n^{\prime\prime}$, $k$, $k^{\prime}$, and $k^{\prime\prime}$ are from $0$ to $N-1$ while $m$, $m^{\prime}$, $m^{\prime\prime}$, $l$, $l^{\prime}$, and $l^{\prime\prime}$ are from $0$ to $M-1$.}
  \begin{equation}\label{ideal_Y}
    Y_{\text{DD}}[k,l]=\sum_{k^{\prime},l^{\prime}}X_{\text{DD}}[k^{\prime},l^{\prime}]h_w[k-k^{\prime},l-l^{\prime}]+Z_{\text{DD}}[k,l]\text{,}
  \end{equation}
  where the term $h_{\omega}$ is the effective channel in DD domain, whose expression can be found in (21) of \cite{OTFS_Window_Design}. By performing the 2D correlation via (\ref{ideal_Y}), we have
    \begin{align}\label{complicated_V}
      &V[k,l]  =\sum_{n,m,k^{\prime},l^{\prime}}X_{\text{DD}}[n-k,m-l]X_{\text{DD}}^{*}[k^{\prime},l^{\prime}]\\
      &\times h^{*}_{\omega}[n-k^{\prime}, m-l^{\prime}]+\sum_{n,m}X_{\text{DD}}[n-k,m-l]Z_{\text{DD}}^{*}[n,m].\nonumber
    \end{align}
  Taking the expectation $V[k,l]$ yields
    \begin{align}\label{E_v_kl}
      &\mathbb{E}[V[k,l]]=\sum_{n,m,k^{\prime},l^{\prime}}\mathbb{E}[X_{\text{DD}}^{*}[k^{\prime},l^{\prime}]X_{\text{DD}}[n-k,m-l]]\\
      &h^{*}_{\omega}[n-k^{\prime}, m-l^{\prime}]+ \sum_{n,m}\mathbb{E}[X_{\text{DD}}[n-k,m-l]Z_{\text{DD}}^{*}[n,m]].\nonumber
    \end{align}
  Since the entries of $\mathbf{X_{\text{DD}}}$ are independent QPSK symbols with unit power, the expectation $\mathbb{E}[X_{\text{DD}}^{*}[k^{\prime},l^{\prime}]X_{\text{DD}}[n-k,m-l]]$ equals $0$ unless $k^{\prime}=n-k$ and $l^{\prime}=m-l$. Meanwhile, the term $\mathbb{E}[X_{\text{DD}}[n-k,m-l]Z_{\text{DD}}^{*}[n,m]]=0$ since the information symbols are independent from the noise samples. Thus, \eqref{E_v_kl} can be simplified as 
  \begin{equation}
    \mathbb{E}[V[k,l]]=MN\cdot h^{*}_{\omega}[k,l]\text{.}
  \end{equation}
  The variance of the entry in matrix $\mathbf{V}$ is
  \begin{equation}
    \text{var}[V[k,l]]=\mathbb{E}[V[k,l]^2]-\mathbb{E}[V[k,l]]^2
  \end{equation}
  where $\mathbb{E}[V[k,l]^2]$ is given in (\ref{v2_complicated}). As before, only the terms with $k^{\prime}=n^{\prime}-k$, $l^{\prime}=m^{\prime}-l$, $k^{\prime\prime}=n^{\prime\prime}-k$, and $l^{\prime\prime}=m^{\prime\prime}-l$ are non zeros. Thus, the second and the third terms on the right-hand side of (\ref{v2_complicated}) can be discarded while the first and last terms are $(MN\cdot h_{\omega}^{*}[k,l])^2$ and $MN\cdot \sigma^2$, respectively.

Consequently, we have
\setcounter{equation}{21}
  \begin{equation}
      \mathbb{E}[V[k,l]^2]=(MN\cdot h_{\omega}^{*}[k,l])^2+MN\cdot \sigma^2\text{,}
  \end{equation}
  which gives the variance of $V[k,l]$, i.e., $\text{var}[V[k,l]]=MN\cdot \sigma^2$. Now, let us consider the variance of $\frac{V[k,l]}{MN}$. Taking the limitation of $\text{var}[\frac{1}{MN}V[k,l]]$ gives
  \begin{equation}\label{limit_v}
    \lim_{M,N\rightarrow\infty}\text{var}\left[\frac{V[k,l]}{MN}\right]=\lim_{M,N\rightarrow\infty}\left(\frac{\sigma^2}{MN}\right)=0\text{.}
  \end{equation}
indicating that when $M$ and $N$ are sufficiently large, the variance of $\frac{1}{MN}V[k,l]$ vanishes. This motivates us to use $\frac{1}{MN}V[k,l]$ to approximate its expectation, i.e. $h_{\omega}^{*}[k,l]$, with an approximation error of $\mathcal{O}\left(\frac{1}{MN}\right)$.

Therefore, the ratio between the magnitudes of the correlation coefficients with the same delay index and Doppler indices $k_{\nu_{1}}$ and $k_{\nu_{2}}$ can be expressed as 
  \begin{equation}
    \begin{aligned}
       & \frac{|V[k_{\nu_{1}}^{\prime},l_{i}]|}{|V[k_{\nu_{2}}^{\prime},l_{i}]|}=\frac{|h_{\omega}[k_{\nu_{1}}^{\prime},l_{i}]|}{|h_{\omega}[k_{\nu_{2}}^{\prime},l_{i}]|}\\
       & =\left\lvert \frac{\sin(-\kappa_{\nu_{i}}\pi)}{\sin(\frac{-\kappa_{\nu(i)}\pi}{N})}\right\rvert \cdot \left\lvert \frac{\sin((k_{\nu_{2}}^{\prime}-k_{\nu_{1}}^{\prime}-\kappa_{\nu_{i}})\pi)}{\sin(\frac{k_{\nu_{2}}^{\prime}-k_{\nu_{1}}^{\prime}-\kappa_{\nu_{i}}}{N}\pi)}\right\rvert ^{-1} \\
       & =\left\lvert \frac{\sin(\frac{k_{\nu_{2}}^{\prime}-k_{\nu_{v1}}^{\prime}-\kappa_{\nu_{i}}}{N}\pi)}{\sin(\frac{-\kappa_{\nu_{i}}\pi}{N})}\right\rvert \approx \frac{|k_{\nu_{2}}^{\prime}-k_{\nu_{1}}^{\prime}-\kappa_{\nu_{i}}|}{|-\kappa_{\nu_{i}}|}\text{.}
    \end{aligned}
  \end{equation}
  Note that the small-angle approximation $\sin x \approx x$ also holds for a sufficiently large $N$. Finally, we arrive at \eqref{derivation}.
\end{appendices}
\bibliographystyle{IEEEtran}
\bibliography{ref}
\end{document}